# A white laser with chirped photonic crystals

A.H. Gevorgyan, N.A. Vanyushkin, I.M. Efimov

School of Natural Sciences, Far Eastern Federal University, 10 Ajax Bay, Russky Island, 690922

Vladivostok, Russia

**We report the results on theoretical investigation about possible simultaneous laser generation in all, closely spaced modes of chirped photonic crystals layer in all visible region of the spectrum. With the possible ability to control the chirping contrast, it allows to obtain a white laser with a simple system on a chip, which can find wide application.**

Multi-color (multi-wavelength) lasers over the full visible-color spectrum from a single chip device have been a subject of great interest in recent years [1–7]. Their ultimate forms are the white lasers [5-12], which can find applications in modern science and techniques, in laser lighting, in full-color laser imaging and display, in biological and chemical sensing, scanning, and monitoring, in ultra-fast data transfer. According to the chromatic diagram, the simultaneous emission of red, green and blue (RGB) can form white lasers, and furthermore, to adjust the chromaticity or improve the color rendering quality, yellow must be added to form the simultaneous emission of red, yellow, green and blue (RYGB), which can also create a white laser [7-9]. The design of such laser devices remains challenging due to the material requirements for producing multi-color output and the complex design for generating laser action, to the balance performance for wide bandwidth, to the high average and peak power, to the high pulse energy, to the high spatial and temporal coherence, and to the profile-driven spectrum. And work in this direction is intensively continuing. Thus, the creation of simple, easily controllable systems on a chip with the possibility of simultaneous laser generation on many modes in the visible region of the spectrum is very relevant. Recently, studies of the possibilities of multi-wave laser generation in disordered systems, and especially white random lasers have been of great interest, too [13,14]. In [15] the authors reported about scheme of harnessing the synergic action of both the second-order nonlinearity and the third-order nonlinearity in a single chirped periodically poled lithium niobate nonlinear photonic crystal driven by a high-peak-power near-infrared (central wavelength~1400 nm, energy~100 µJ per pulse) femtosecond pump laser to produce visible to near infrared (400-900 nm) supercontinuum white laser.

On the other hand, for multimode laser generation in a wide frequency region, and on many modes one can use a chirped, distributed Bragg reflector (DBR) photonic crystal (PC) structure. Especially if we keep in mind that in this case there is a parameter that in principle can be controlled, namely the chirping contrast. In this paper we will theoretically investigate the possibilities of simultaneous multiwavelength laser generation by chirped PCs at different values of contrast of chirping.

We will consider a chirped 1D PC of the form

$$\varepsilon(z) = \varepsilon + \Delta\varepsilon \sin\left(\frac{2\pi}{\Lambda(z)} z\right) \quad (1)$$

where $\varepsilon$ is the constant value, $\Delta\varepsilon$ and $\Lambda(z)$ are the depth and the period of modulation, correspondingly. The PC layer with thickness $d$ is sandwiched between the planes $z=0$ and $z=d$, and $\varepsilon_s = 1$, where $\varepsilon_s = n_s^2$ is the dielectric permittivity of the medium, bordering with the PC layer on both sides. We will consider the linear law of chirping, moreover, both with positive and negative slope, namely, the first case with $\Lambda(z) = az + b = \frac{\Lambda_{max} - \Lambda_{min}}{d} z + \Lambda_{min}$, with $a = \frac{\Lambda_{max} - \Lambda_{min}}{d}$, and $b = \Lambda_{min}$. In the second case we will have $a = \frac{\Lambda_{min} - \Lambda_{max}}{d}$ and $b = \Lambda_{max}$. In the first case, along the direction of light propagation, the modulation period increases linearly from the value of $\Lambda_{min}$ at the input surface to the value of $\Lambda_{max}$ at the output surface, and in the second case, it decreases linearly from the value of $\Lambda_{max}$ at the input surface to the value of $\Lambda_{min}$ at the output surface. The method for solving this problem is described in all details in the paper [16].

Here we consider the light normal incidence case only. We will investigate the peculiarities of light localization, specific properties of spectra of reflection, of absorption, of photonic density of states (PDOS) and of density of light energy. The electric field intensities in a 1D PC layer are defined as:
$$I_{in}(z) = |E_{in}(z)|^2, \qquad (2)$$
and the absorption coefficients as: $A = 1 - (T + R)$. Here $E_{in}(z)$ is the electric field inside the PC layer, $T$ and $R$ are the transmission and reflection coefficients, correspondingly. The light energy density in the PC layer will be calculated by the following formula:
$$w = \frac{1}{d}\int_0^d |E_{in}(z)|^2 dz, \qquad (3)$$
and the PDOS by the following formula [17]:
$$\rho_l = -\frac{\lambda^2}{2\pi c}\frac{dk_l}{d\lambda} = -\frac{\lambda^2}{2\pi c d}\frac{v_l\frac{du_l}{d\lambda} - u_l\frac{dv_l}{d\lambda}}{u_l^2 + v_l^2}, \qquad (4)$$
where $u_l$ and $v_l$ are the real and imaginary parts of the amplitude of the transmitted wave, the values $l = 1, 2$ correspond to the eigenmodes of the PC layer (p- and s- modes), respectively. We will investigate these dependences at different values of contrast of chirping $C = \Lambda_{max} - \Lambda_{min}$. First of all, let us note, that as showed in [16, 18,19] the gradient of the period of modulation has influence on the photonic band gap (PBG) widening, and for PBG boundaries we have $\lambda_1 = 2n_o\Lambda_{min}$ and $\lambda_2 = 2(n_e\Lambda_{max} + \bar{n}\frac{d\Lambda}{dz}\Delta z)$ in the first case of chirping, while for the case of the second type chirping they are defined as: $\lambda_1 = 2(n_o\Lambda_{min} + \bar{n}\frac{d\Lambda}{dz}\Delta z)$ and $\lambda_2 = 2n_e\Lambda_{max}$. Here $n_{o,e} = \sqrt{\varepsilon \pm \frac{\Delta\varepsilon}{2}}$ and $\bar{n} = \frac{n_o + n_e}{2}$.

Fig.1(a) shows the schematics of the problem: planar PC layer. The rest of figures are the evolution of the light intensity inside the PC layer $I(z) = |E_{in}(z)|^2$ when the coordinate $z$ changes in the case of the absence of chirping (b), at contrast of chirping $C = 40$ nm (c,d), at $C = 100$ nm (e,f), and finally at $C = 200$ nm (g,h). Fig. 1 (c,e,g) corresponds to the case of positive slope of chirping, and Fig. 1 (d,f,h) to the case of negative slope.

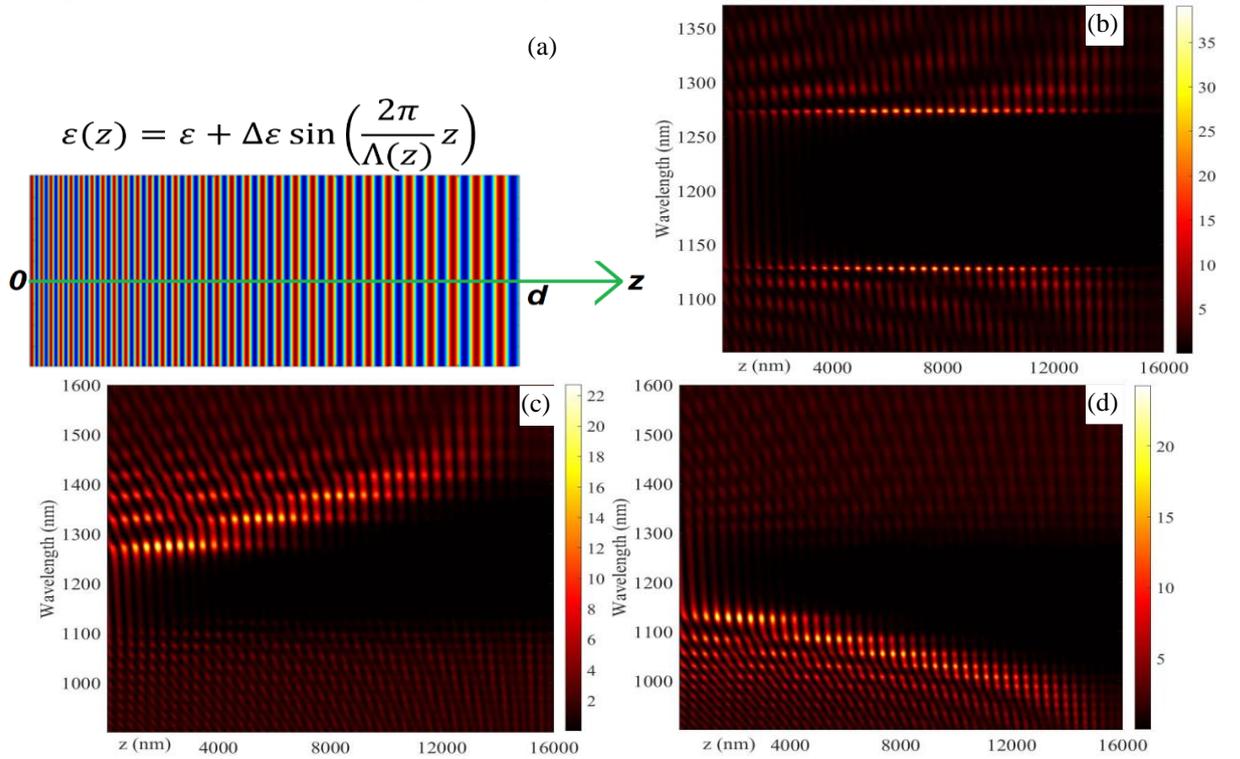

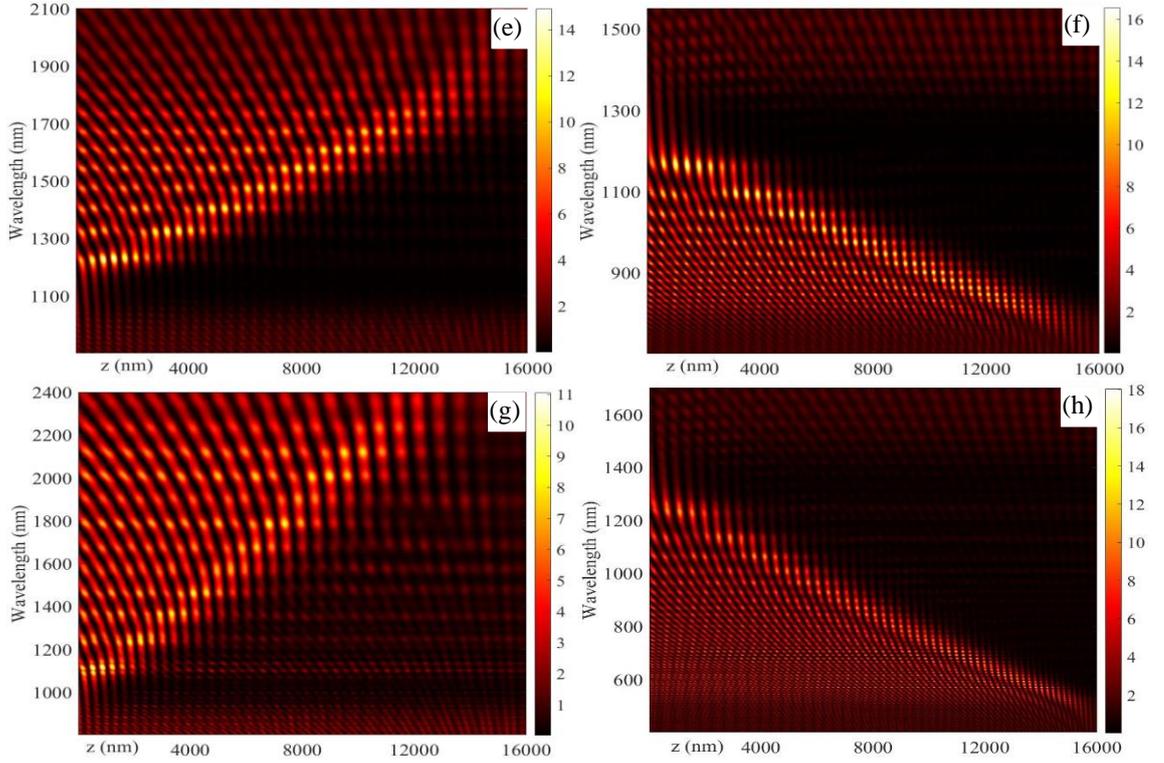

Fig.1. (a) Schematics of the problem: planar PC layer. The evolution of the light intensity inside the PC layer $I(z) = |E_{in}(z)|^2$ when the coordinate $z$ changes in the case of the absence of chirping (b), at contrast of chirping $C = 40$ nm (c,d), at $C = 100$ nm (e,f), and finally at $C = 200$ nm (g,h). Fig. 1 (c,e,g) corresponds to the case of positive slope of chirping, and Fig. 1 (d,f,h) to the case negative slope. The parameters of PC layer are: $\varepsilon = 2.25$; $\Delta\varepsilon = 0.5$; $d = 16$ μm; $\Lambda_{min} = \Lambda - \frac{C}{2}$; $\Lambda_{max} = \Lambda + \frac{C}{2}$; $n_s = 1$. $\Lambda = 400$ nm and it is the period of modulation in the chirping absence case. The absorption is absent.

Fig. 2 shows the spectra of $R$ reflection (curve 1), $A$ absorption (curve 2), $w$ light energy density (curve 3) and of $\rho_l/\rho_{iso}$ relative PDOS (curve 3), again in the case of the absence of chirping (a), at contrast of chirping $C = 40$ nm (b,c), at $C = 100$ nm (d,e), and finally at $C = 200$ nm (f,g). Fig. 2 (b,d,f) corresponds to the case of positive slope of chirping, and Fig. 2 c,e,g to the case negative slope. $\rho_{iso} = n_s/c$, where $c$ is the speed of light in a vacuum.

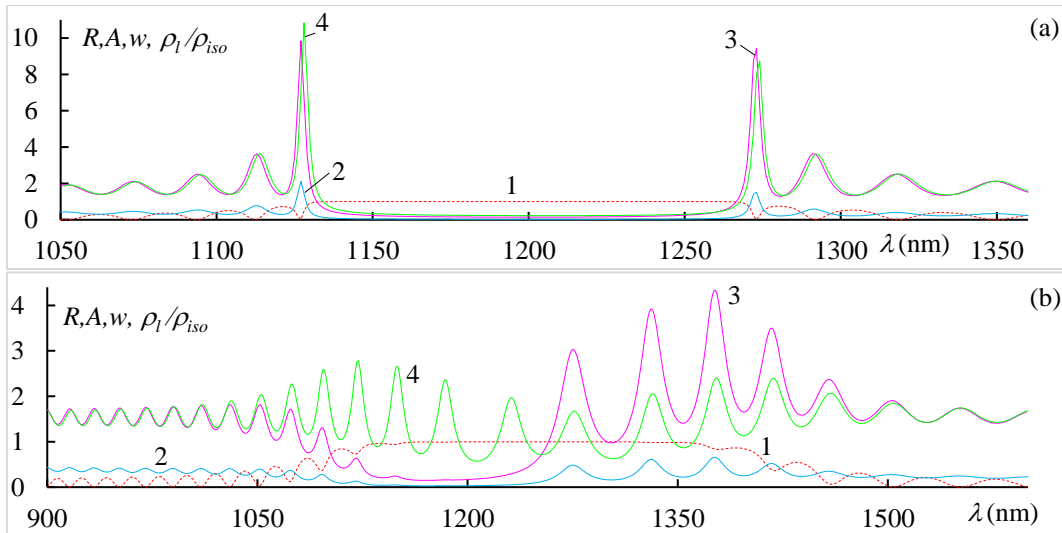

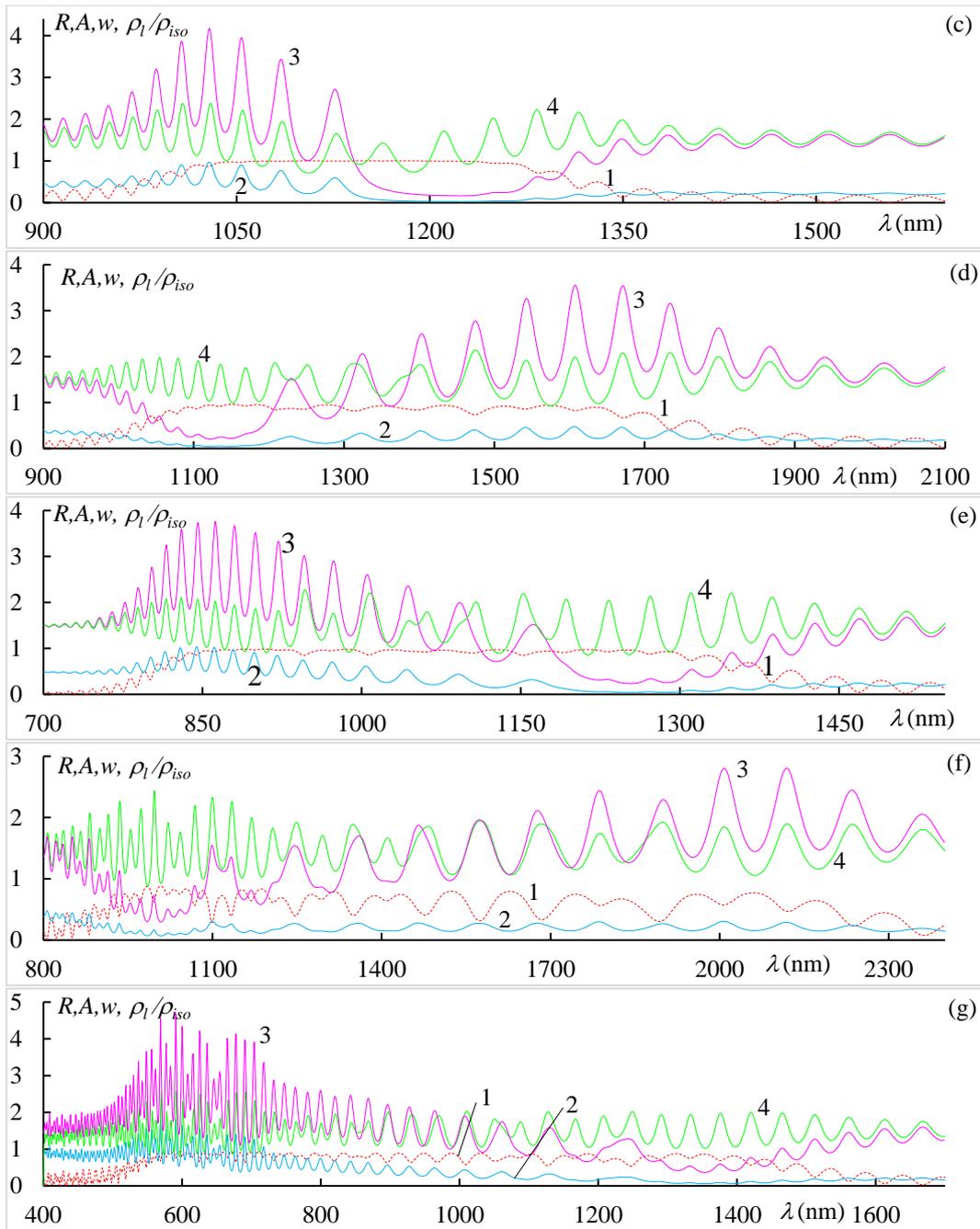

Fig.2. The spectra of $R$ reflection (curve 1), $A$ absorption (curve 2), $w$ light energy density (curve 3) and of $\rho_l/\rho_{iso}$ relative PDOS (curve 3), again in the case of the absence of chirping (a), at contrast of chirping $C = 40$ nm (b,c), at $C = 100$ nm (d,e), and finally at $C = 200$ nm (f,g). Fig. 2 (b,d,f) corresponds to the case of positive slope of chirping, and Fig. 2 (c,e,g) to the case negative slope. The absorption is homogeneous and $\text{Im}\varepsilon = 10^{-5}$. The other parameters are the same is in Fig.1.

As follows from Figures 1 and 2 in the case of chirping, the widening of PBG takes please and with increase of contrast $C$ this widening increases, too, and the dips in the reflection spectrum appear (Zener tunneling [16,20]). Moreover, there appear coupled modes due to Bloch oscillations in chirped crystals and where strong light localization takes place and low threshold lasing is possible. With increase of contrast $C$ the density of coupled modes increases. So, in chirped PCs unlike to PCs with ideal periodical structure two type of modes are excited, the edge modes and the coupled ones.

Now we pass to investigate the thresholds of laser generation at these coupled modes as well as at edge modes. We will consider our PC to reach the lasing threshold when the denominator of the transmission coefficient *t* turns into zero, i.e. the transmission reaches infinity [21]. This condition can be interpreted as when lasing is achieved one can see a finite outcoming wave in the absence of any incoming wave. So, if the condition $1/t(\lambda, \text{Im}\varepsilon) = 0$ is satisfied for some pair of parameters $(\lambda, \text{Im}\varepsilon)$, where $\text{Im}\varepsilon < 0$ is the amplification, then the lasing is achieved at the wavelength $\lambda$ with the lasing threshold $\text{Im}\varepsilon_{th}$. It is also worth noting that the lasing can be achieved only at specific wavelengths while at all other wavelengths the lasing is impossible no matter how much the amplification is.

Fig.3 shows the values of $|\text{Im}\varepsilon_{th}|$ for lasing modes (a) in the case of the absence of chirping, and (b,c) at contrast of chirping $C = 200$ nm. Fig. 3 (b) corresponds to the case of positive slope of chirping, and Fig. 3 (c) to the case of negative slope. As can be seen from Fig. 3 (c) at value $|\text{Im}\varepsilon|$=0.02 simultaneous laser generation in all, closely spaced modes in all visible region of the spectrum is possible. Then, at $|\text{Im}\varepsilon|$=0.08 simultaneous laser generation in all modes in the spectral region from 400 nm to 2400 nm is possible, moreover both in forward and backward directions.

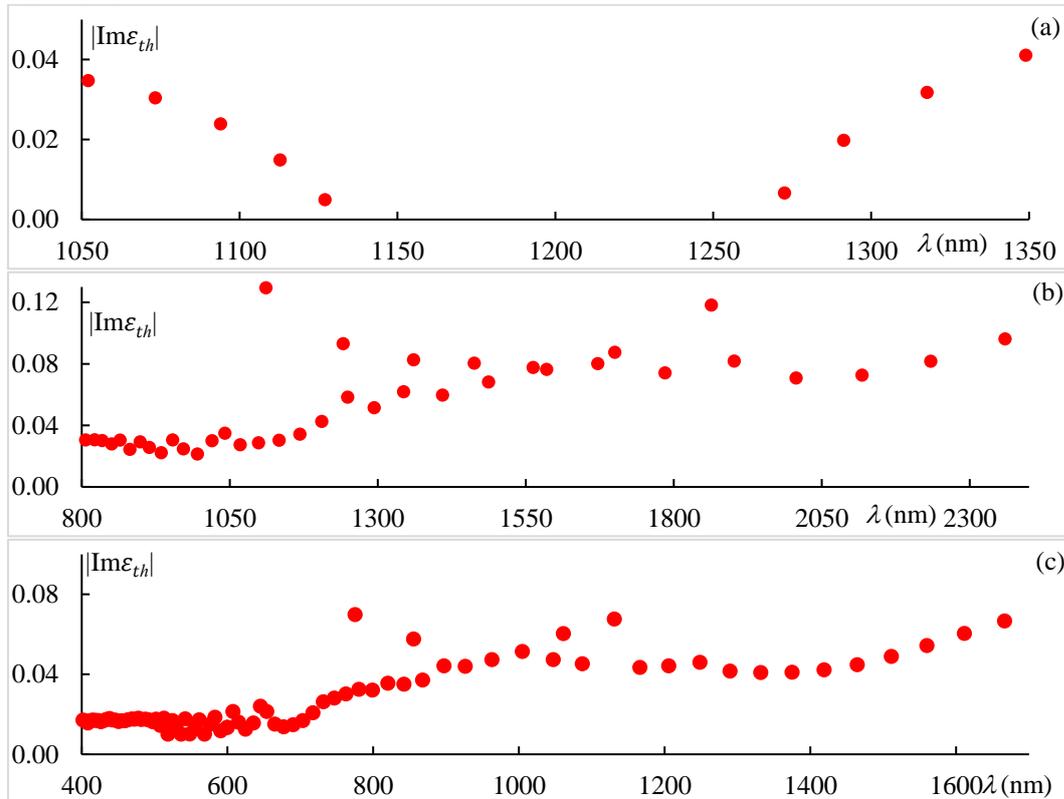

Fig.3. The values of $|\text{Im}\varepsilon_{th}|$ for lasing modes (a) in the case of the absence of chirping, and (b,c) at contrast of chirping $C = 200$ nm. Fig. 3 (b) corresponds to the case of positive slope of chirping, and Fig. 3 (c) to the case of negative slope. The other parameters are the same is in Fig.1.

In conclusion, we theoretically investigated the possible simultaneous laser generation in all, closely spaced modes of chirped PCs layer in the entire visible region of the spectrum. In addition, we note that with the ability to control the chirp contrast, it becomes possible to obtain a white laser with a simple system on a chip, which can find wide application.

**Disclosures:** The authors declare no conflicts of interest.